\documentclass[aps,prl,twocolumn,superscriptaddress,showpacs]{revtex4}

\usepackage{amsmath,amssymb,graphicx}
\def\ket#1{\mathinner{|{#1}\rangle}}
\def\bra#1{\mathinner{\langle{#1}|}}
\def\ket#1{\mathinner{|{#1}\rangle}}

{\catcode`\|=\active
  \gdef\Braket#1{\begingroup
     \ifx\SavedDoubleVert\relax
       \let\SavedDoubleVert\|\let\|\BraDoubleVert
     \fi
     \mathcode`\|32768\let|\BraVert
     \left<{#1}\right>\endgroup}
}
\def\BraVert{\@ifnextchar|{\|\@gobble}
     {\egroup\,\mid@vertical\,\bgroup}}
\def\BraDoubleVert{\egroup\,\mid@dblvertical\,\bgroup}
\let\SavedDoubleVert\relax
\newcommand{\vectornorm}[1]{\left|\left|#1\right|\right|}

\begin{document}
\title{Fractal Weyl law behavior in an open, chaotic Hamiltonian system}

\author{Jordan A. Ramilowski}
\affiliation{Departamento de Qu\'{\i}mica, and
Instituto Mixto de Ciencias Matem\'aticas CSIC--UAM--UC3M-UCM,
Universidad Aut\'onoma de Madrid, Cantoblanco, 28049 Madrid, Spain}
\affiliation{Department of Chemistry, Utah State University,
Logan, UT 84322-0300}

\author{S. D. Prado}
\affiliation{Departamento de Qu\'{\i}mica, and
Instituto Mixto de Ciencias Matem\'aticas CSIC--UAM--UC3M-UCM,
Universidad Aut\'onoma de Madrid, Cantoblanco, 28049 Madrid, Spain}
\affiliation{Instituto de F\'{\i}sica,
Universidade Federal do Rio Grande do Sul, PO Box 15051,
91501--970 Porto Alegre, RS, Brazil.}

\author{F. Borondo}
\affiliation{Departamento de Qu\'{\i}mica, and
Instituto Mixto de Ciencias Matem\'aticas CSIC--UAM--UC3M-UCM,
Universidad Aut\'onoma de Madrid, Cantoblanco, 28049 Madrid, Spain}

\author{David Farrelly}
\affiliation{Departamento de Qu\'{\i}mica, and
Instituto Mixto de Ciencias Matem\'aticas CSIC--UAM--UC3M-UCM,
Universidad Aut\'onoma de Madrid, Cantoblanco, 28049 Madrid, Spain}
\affiliation{Department of Chemistry, Utah State University,
Logan, UT 84322-0300}

\date{\today}
\begin{abstract}
We numerically show fractal Weyl law behavior in an open Hamiltonian
system that is described by a smooth potential and which supports
numerous above-barrier resonances.
This behavior holds even relatively far away from the classical limit.
The complex resonance wave functions are found to be localized on the
fractal classical repeller.
\end{abstract}
\maketitle
%
The classical and quantum dynamics of open Hamiltonian systems is relevant
to a variety of topics of current interest in macroscopic and microscopic physics.
For example, in planetary physics, the formation of binaries in the Kuiper-belt
may have proceeded through the formation of transitory objects in chaotic layers
of phase space trapped close to above-barrier Kolmogorov-Arnold-Moser (KAM)
islands \cite{Perry94,Asta03,Lee07}.
The analogs of these states in open quantum Hamiltonians are resonances (quasibound states)
which, in general, are predicted to
be localized on an object known as the classical repeller \cite{Gaspard:1998}.
The repeller is the intersection of two fractal sets of classical trajectories
one of which remains trapped in the infinite past and the other in the infinite future,
denoted as $K_-$ and $K_+$, respectively.
The fractal nature of these sets has led to the prediction of a fractal Weyl law
for flows in which the number of long-living quantum resonances scales as
$\hbar^{-(1+d_H)}$ where $d_H$ is the partial Hausdorff dimension of the repeller
\cite{Lu03}.

In open maps the number of resonances has already been found to obey a
similar fractal Weyl law, $\hbar^{-d}$,
except that $d$ is now the partial fractal dimension of the trapped set.
This relates to the original Weyl law \cite{Baltes78} conceived for closed
systems which states that the number of eigenstates up to energy E
which fits into the available phase-space volume of the classical system
scales as $\hbar^{-d}$, with $d$ being the actual (integer) dimensionality
of quantum space.
In addition, in open maps the associated quantum Husimi distributions are observed to
cling to the classical repeller \cite{Prado06,Shepelyansky:2008}.
Although the fractal Weyl law has been observed in particular maps like the baker
map and the kicked rotor
\cite{Lu03,Sch2004,Prado06,Nonnenmacher:2006,Shepelyansky:2008,Keat2008,Ermann:2009},
there have been few previous studies of this problem in open
{\it Hamiltonian} systems \cite{Lin02} even though such systems are of direct physical interest;
e.g., the chaotic ionization of hydrogen atom interacting with a circularly
polarized microwave exhibits above-barrier chaotic trapping \cite{Brunello97}.

Here we present an examination of above-barrier quantum resonance (Gamow) states in
a model Hamiltonian whose classical dynamics is chaotic. The system is described
by a smooth potential and numerous above-barrier resonances are supported.
The model we use is chosen to capture essential features of the chaotic ionization
dynamics of atoms in rotating fields;
further, we propose, that this mechanism may also be important in complex formation
in certain chemical reactions.
We find that, not only does the fractal Weyl law hold for a typical (generic)
open Hamiltonian system \cite{Kopp2009}, but it also holds even in the vicinity
of $\hbar =1$, i.e., far from the classical limit.

The investigation of a fractal law for open analytical Hamiltonian
systems is problematic for a number of technical reasons,
including the larger dimensionality of phase space $N$ needed to observe the chaotic repeller;
unlike in unidimensional maps, for which $N = 2$, in an autonomous Hamiltonian the repeller
exists only if $N \ge 4$ which compounds the computational challenges involved.
An additional computational difficulty is the calculation of resonance eigenfunctions in
the limit $\hbar \rightarrow 0$ because of the attendant growth in the size of the basis
needed to converge the calculations.
Some of these obstacles have been overcome in a previous study which reported fractal Weyl law
behavior in an open Hamiltonian whose potential energy surface (PES) consisted of three
gaussian bumps \cite{Lin02}.
However, for computational reasons only a rather limited number of resonances were included
in the analysis and the structure of the resonance eigenstates themselves was not examined.

The model chosen here provides a realistic model of the chaotic ionization of atoms and
of resonances in chemical reactions in that the PES features a potential well together
with saddle points and, depending on the energy, the classical dynamics may be mixed
(i.e., regular and chaotic) even above the saddle points.
Computation of quantum complex (resonance) eigenvalues and quantum surfaces of section
(QSOS, see, e.g., Ref.\  \cite{Ermann:2009, Dando94}), based on Husimi distributions,
reveals that the above-barrier resonance energies are localized on the classical repeller.
For values of $\hbar$ away from the asymptotic limit there is progressively more
delocalization.
The advantage of the model used is that the calculations are more tractable than for,
say, the H atom interacting with rotating fields for which the Coulomb term complicates
the computations.

The model is a modification of the H{\'e}non-Heiles (HH) Hamiltonian \cite{Kaidel04}
%
\begin{equation}
  H=
 \frac{1}{2}(p_x^2+p_y^2)+\frac{1}{2}(x^2+y^2) +
  \lambda (x^2 y - \frac{1}{3}y^3) - \omega \thinspace ( x p_y - y p_x)
     \label{eq:Hamiltonian}
\end{equation}
where, throughout, $\lambda = 0.1$ and $\omega = 0.1$.
The modification is the presence of a Coriolis term - the term in $\omega$ -
which is designed to simulate the addition of,
e.g., a CPM field or a magnetic field to a Rydberg atom \cite{Buch2002}.
Because the Hamiltonian does not have rotational symmetry this
angular-momentum-like term is not a conserved quantity.
Furthermore, the presence of the Coriolis term means that it is no longer
possible to define a potential energy surface - instead one can resort to using
the device of a zero velocity surface (or ZVS, see, e.g., Ref.~\cite{Brunello97}).
Finally, time reversal symmetry of the system is broken.
Energies and widths are scaled by the energy of the three saddle points in the ZVS,
i.e., $E_s = (1-\omega^2)^3/6\lambda^2 = 16.17165$.

\begin{figure}
\includegraphics[width=5cm,angle=270]{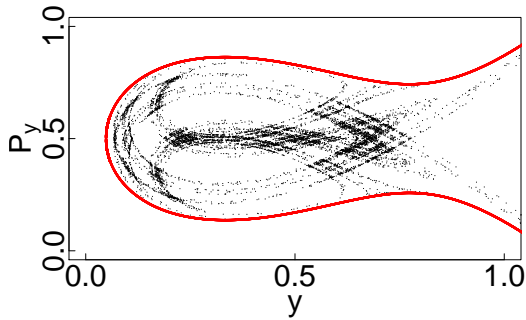}
 \caption{(Color online)
   Superposition of the two branches of the repeller computed as described in the
   text and projected onto the SOS defined by $x=0, \dot{x} < 0$ and $E = 1.8 E_s$.
   The bounding curve, solid line (red online), of the SOS is also shown.
   Both axes have been scaled to the interval (0,1) to allow for comparison with
   the Husimi plot of Fig.~\ref{fig:3}.}
    \label{fig:1}
\end{figure}

The structure of the repeller is shown in Fig.~\ref{fig:1} for $\omega = 0.1$
and an energy rather high above the saddle point energy.
The repeller was computed by integrating trajectories forwards and backwards in time.
The initial conditions of trajectories which survived (i.e., did not escape)
for a time $\tau_0$ were saved.
Survivors were then reintegrated for a time $20 \tau_0$ and their intersections
with an appropriate Poincar{\'e} surface of section (SOS) were recorded.
In this case the SOS chosen is defined by $x=0, \dot{x} < 0$.

The effect of adding the Coriolis term is that, for nonzero $\omega$,
relatively large KAM islands may co-exist with - but are not part of -
the repeller for energies considerably above the energy of the saddle points.
Here we do not consider resonances directly associated with these islands.
However, the Coriolis terms has the effect of bringing out the structure of the
repeller more clearly than for the pure HH system ($\omega = 0$ \cite{Kaidel04}).
Thus, varying $\omega$ allows for the fine tuning of the dynamics in the energy
regime of interest.

The method of complex rotation was used to compute the complex resonance energies
$E_n = E_r - i \Gamma_n/2$ where $\Gamma_n$ is the resonance width \cite{Ho83}.
This was accomplished by rotating the coordinates into the complex plane by an
angle $\theta$, i.e., $q_i \rightarrow q_i e^{i \theta}$ and then diagonalizing
the resulting Hamiltonian matrix in a two-dimensional isotropic oscillator basis
$\ket{n,m}$ \cite{Kaidel04}.
In principle the procedure is straightforward although care must be exercised to
ensure that resonances are distinguished from scattering states.
This can be accomplished by examining so-called $\theta$-trajectories, i.e.,
as the angle $\theta$ is varied the resonances, as distinct from scattering states,
converge.
A large number of resonances lying above the saddle points are required to achieve the
quality of statistics needed to determine how the number of resonances scales with $\hbar$.

The complex energy spectrum for the Hamiltonian of eq.~\ref{eq:Hamiltonian} contains
resonances lying below as well as above the saddle points.
Sub-saddle states decay by tunneling.
However, the states of interest lie above the saddle points and, therefore,
a large basis set must be used to converge these resonances.
As $\hbar$ is decreased the number of states below the saddles grows and, therefore,
the size of the basis must be increased.
For this reason it is difficult to access the very small values of $\hbar$
- or equivalently, the very high-lying states - for which it is normally assumed
that the fractal Weyl law will hold.
Working on the observation that, in general, asymptotic expansions often provide
good agreement even outside their strict domains of validity, we examined resonance
statistics for $\hbar$ in the vicinity of  $\hbar = 1$.
Numerically, resonances were computed by direct diagonalization and also, as check,
by using the Arnoldi method which takes advantage of the sparsity of the Hamiltonian
matrix \cite{Lin02}.
The Arnoldi method has the twin merits that
(i) a larger basis can be employed, and
(ii) it allows access to selected portions of the spectrum.
However, only a relatively small subset of resonances can be computed in this
way \cite{templates}.

\begin{figure}
\includegraphics[width=5.00cm,angle=270]{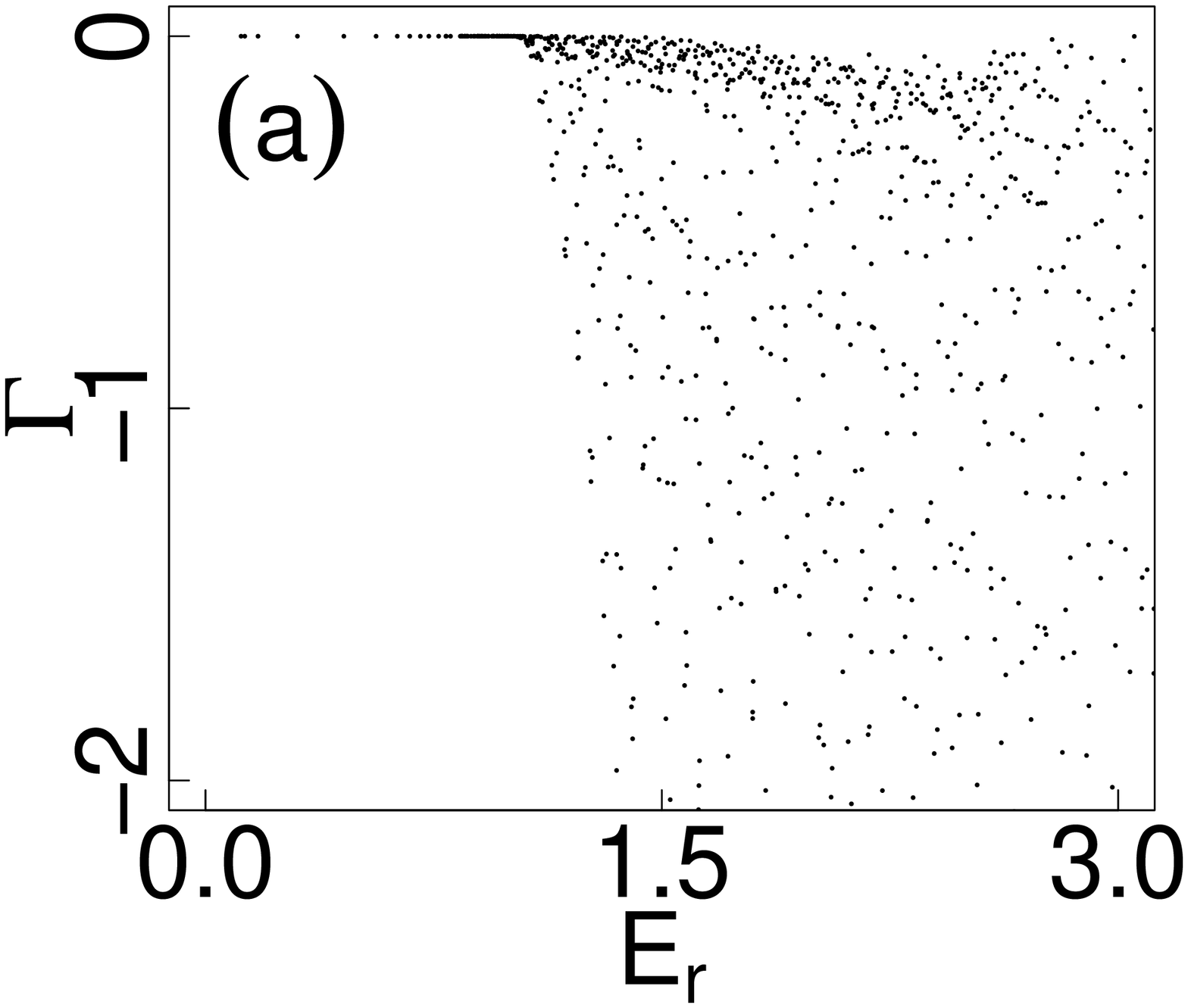}
\includegraphics[width=5.00cm,angle=270]{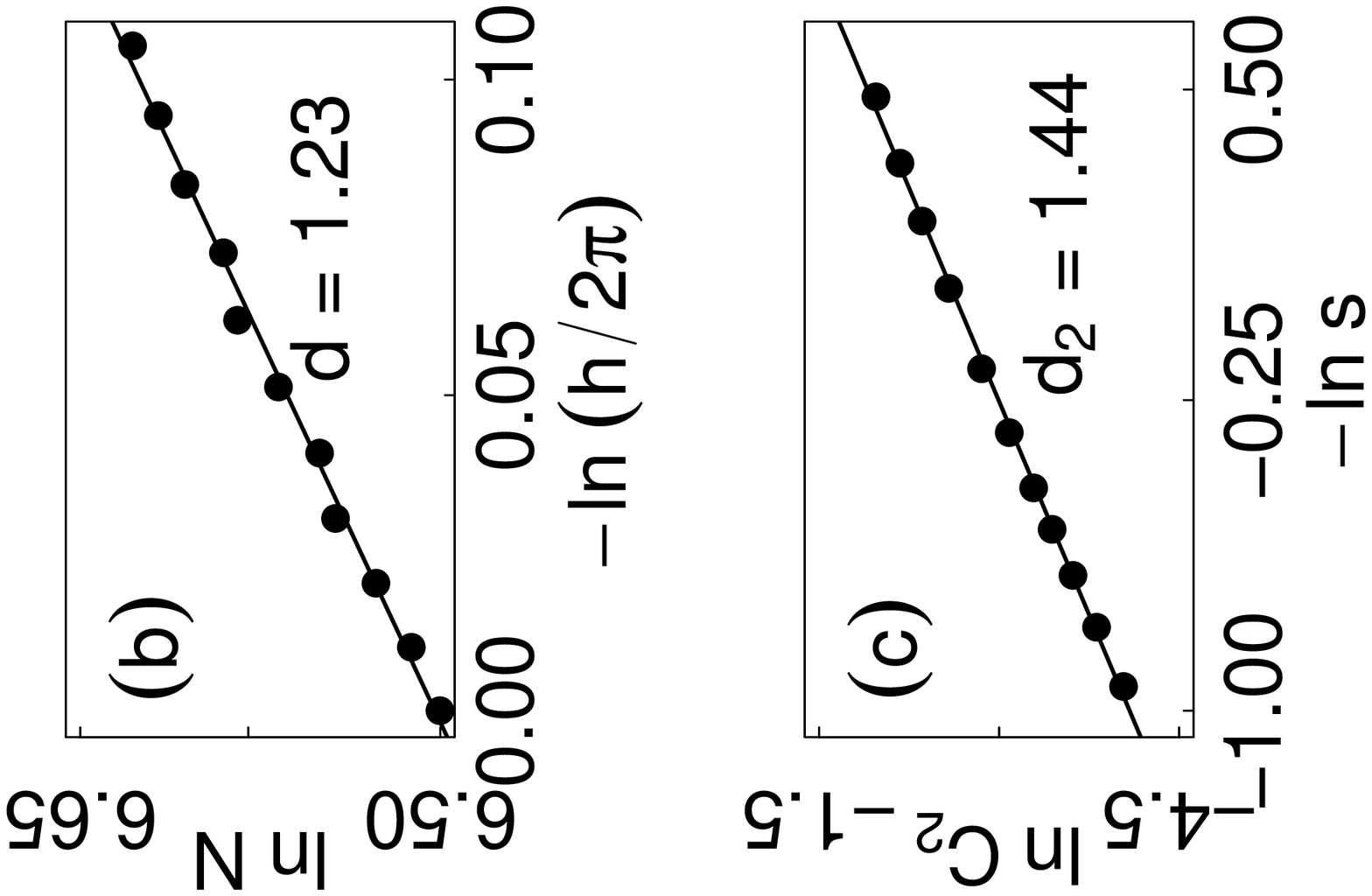}
 \caption{Frame (a) shows resonance positions ($E_r$) and widths ($\Gamma$) (scaled by the
 saddle point energy $E_s$) for $\hbar =1$.
 In (b) the classical quantity $\ln C_2(s)$ is shown {\it vs.}
 $\ln s$ together with a best fit to a line whose slope is
 the correlation dimension [see eq.~(\ref{eq:2})], i.e., $d_2 = 1.442\pm 0.008$.
 Frame (c) shows a best fit to box counted quantum resonances
 with slope $d = 1.231 \pm 0.028$.
 The quantum resonances scale as
 $\hbar^{-m/2}$ or $\hbar^{-1.22}$\cite{Lu03} where $m=(1+d_2)$ is the dimension
 of the trapped set;
 the quantum box counting dimension is $1.23$.
 Despite considering a relatively narrow range of $\hbar$,
 lying far from the classical limit the classical and quantum fractal dimensions agree well.}
\label{fig:2}
\end{figure}

Distributions of resonances in the range $0.9 \le \hbar \le 1$ were then computed.
Figure \ref{fig:2} shows the complex resonance eigenvalues obtained for $\hbar=1$.
By counting the number of states, $N(\hbar)$, in 8 different rectangular boxes
of size $(1,1.24\hbar)$ located around $E_r=1.8 E_s$ and
averaging over these sets of data we were able to establish that
the number of states follow a Weyl law with $d = 1.231 \pm 0.028$
(see lower inset in Fig.\ \ref{fig:2} ).
Similarly, and also as shown in the figure, the dimension of the classical
repeller of Fig.~\ref{fig:1} computed from the Poincar{\'e} map is fractal
with correlation dimension $d_2= 1.442 \pm 0.008$ where
\cite{Lauterborn:1988}:
%
\begin{equation}
  d_2=\lim_{s\rightarrow 0}\frac{\ln C_2(s)}{\ln s}.
  \label{eq:2}
\end{equation}
Here $s$ is the edge length of an $n$-dimensional cube and $C_2(s)$ is the correlation
sum \cite{Lauterborn:1988}
%
\begin{equation}
  C_2(s)=-\lim_{M\rightarrow \infty}\frac{1}{M^2}\sum_{k,\ell=1}^M\Theta
\left(s-\vectornorm{\mathbf{q}_k^n-\mathbf{q}_\ell^n}\right)
\end{equation}
where $M$ is the number of points in the repeller, $\Theta$ is the Heaviside step
function and $\mathbf{q}_k$ are the points of the repeller.
The fractal dimension, $m$, is related to the correlation dimension computed from
a Poincar{\'e} map as $m=1+d_2$ \cite{Lauterborn:1988}.
Using the classical data we found that $d_2 = 1.442 \pm 0.008$ which leads to a
fractal dimension for the repeller of $m=2.44$.

According to Ref.\ \cite{Lu03} the quantum resonances in an energy interval
should scale as $\hbar^{-m/2}$ where $m$ is the dimension of the trapped
set for the energies in that interval.
This is in excellent agreement with quantum box counting since $m/2=1.22$
while the quantum box counting gives $d=1.23$.
This prediction is borne out remarkably well by Fig.~\ref{fig:2} and is an illustration
of the fractal Weyl law in an open Hamiltonian system (rather than in a map).
The number of resonances in the boxes varied from 686 to 827 depending on $\hbar$.
%

In addition to the  conjecture that for generic open systems fractal
Weyl law behavior will be observed, it is also expected that Husimi
functions will coagulate onto fractal sets (i.e., onto the repeller)
in the limit $\hbar \rightarrow 0$ \cite{Prado06}.
For finite $\hbar$ Husimi functions will not truly be confined to fractal sets
and will appear somewhat blurred due to quantum effects.
However, as $\hbar$ is decreased then classical structures, on progressively finer scales,
will become apparent in the Husimi functions.
The $\hbar \rightarrow 0$ limit is itself of physical interest in that this limit
corresponds, e.g., to the ionization of ultrahigh Rydberg states.

Because the Hamiltonian is complex and non-Hermitian the left, $\Psi_L^{(i)}$,
and right, $\Psi_R^{(i)}$ eigenfunctions do not satisfy the usual
(Hermitian) identity $\Psi_L^{(i)} = \Psi_R^{(i)*}$ and, consequently,
$\rho_i = \Psi_L^{(i)} \Psi_R^{(i)*}$ is a complex quantity.
In fact, observables are associated with neither $\Psi_L^{(i)}$ nor $\Psi_R^{(i)}$
but with $\sqrt{\Psi_L^{(i)} \Psi_R^{(i)} }$ \cite{nim}.
This complicates the computation of Husimi distributions as has been discussed
by Buchleitner, {\it et al.} \cite{Delande94} who pointed out that Husimi distributions
for individual complex eigenstates have the peculiar property that they can be negative
and a sum needs to be made, in principle, over all complex energies \cite{Delande94}.
For this reason, and, in analogy with previous computations in quantum maps,
Husimi distributions are averaged over an energy range of finite width.
We use the following definition of the averaged Husimi function,
whose derivation includes both left and right eigenstates,
and which is in the spirit of Bogomolny \cite{Prado06,Delande94,Bog1988}
%
\begin{equation}
  |\langle{\Omega}\ket{\phi_E}|^2=\frac{1}{\pi} \mbox{Im}\sum_i
  \frac{\bra{\bar{\Psi}_{L}{^{(i)}}}R(\theta)\ket{\Omega}
  \bra{\bar{\Psi}_{L}{^{(i)}}}R(\theta)\ket{\bar{\Omega}}}{E_{i \theta} -E}.
     \label{eq:Husimi}
\end{equation}
Here $\phi_E$ represents the probability amplitude at real energy $E$ and
$\ket{\Omega}$ is a coherent state; $\Psi_{L}$ is a complex rotated eigenstate
expressible in terms of the isotropic oscillator basis vectors;
$E_{i \theta}$ is the complex energy of the eigenstate and the overbar notation
signifies, e.g., that $\bra{\bar{\Psi}}$ is the complex conjugate of
$\bra{\Psi}$; $R(\theta)$ is the complex rotation operator \cite{Delande94}.
Very recently, Ermann et al. have proposed a different, although related phase space
representation for open quantum systems \cite{Ermann:2009}.

For narrow resonances simplifications of eq. (\ref{eq:Husimi}) are possible.
By projecting the states onto a basis of isotropic oscillator functions one
avoids computing basis vectors in the complex coordinate plane,
a procedure which is numerically unstable since basis vectors which are oscillatory
along the real axis may diverge exponentially in the complex plane \cite{Delande94}.
In this case one then needs to compute matrix elements of $R(\theta)$ in the basis
used although these matrix elements themselves ultimately diverge -
the resonance eigenfunctions are not $\mbox{L}^2$ functions.

We adopted the following procedure to project the 4-dimensional Husimi distribution
onto a 2-dimensional hypersurface in phase space so as to generate a QSOS.
A narrow interval of energy was selected around some energy of interest $E_0$.
Equation (\ref{eq:Husimi}) was then used with the resonance eigenstates projected
onto the isotropic oscillator basis. Only resonance states with widths smaller
than some width, $\Gamma_0$, were included in the summation and the matrix elements
$\bra{n}R(\theta)\ket{m}$ were approximated by their lowest order (i.e., diagonal)
expansion in $\theta$.
For states with narrow widths, as is the case here, this is an excellent approximation.
The QSOS was then computed by fixing $x=0$ and computing the conjugate momentum $p_x$
using the classical Hamiltonian at energy $E$.
Because the boundary of the classically allowed region itself changes with energy
this procedure is not entirely satisfactory when computing an average Husimi QSOS.
However, provided that the energy range is kept sufficiently small the errors so
introduced are expected to be minimal and this was verified by direct computation.

\begin{figure}
\includegraphics[width= 9 cm]{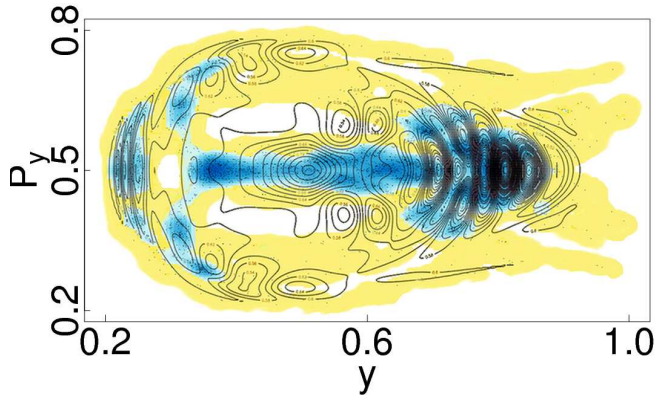}
 \caption{(Color online) Contours of the averaged Husimi function calculated as described
 in the text superimposed on a representation of the classical repeller.
 Twenty resonance states on each side of  $E = 1.8 E_s$ were included.
 The points which represent the classical repeller on the SOS have been
 kernel smoothed which, in essence, assigns a local density of points and
 then colors that section of the plot accordingly \cite{Lee07} -
 compare to Fig.~\ref{fig:1}.
 The color scale runs from white through light grey (yellow and light blue online) to dark grey (dark blue online) and represents the density from low to high accordingly.}
    \label{fig:3}
\end{figure}

Figure 3 shows a typical example of a QSOS computed in this way.
The averaged Husimi distribution is clearly localized on the fractal repeller sets
$K_+$ and $K_-$ which are also shown in the figure.
However, this tendency for the Husimi to coagulate onto the repeller is mitigated
by the relatively large value of $\hbar = 1$ used in constructing the figure.
The Husimi distribution is somewhat delocalized over the repeller and does not
precisely follow the contours of the fine scale classical structures.
It is also apparent in Fig.\ 3 that the quantum density builds up close to the saddle point.
The reason for this is that the quantum particle senses the presence of classical
turning points in the complex plane and, therefore, slows down which leads to a build-up
in probability density in the vicinity of the saddle point.
This is consistent with the recent findings of Keating {\it et al.}
who note that for longer living states, the long lifetime allows interference
and diffraction effects to accumulate thereby washing out the fractal structure
to some extent \cite{Keat2008}.

In summary: the fractal Weyl
law was found to hold in an open Hamiltonian system.
Despite working far from the asymptotic limit
$\hbar \rightarrow 0$
the resonance energies manifested clear fractal behavior and averaged
Husimi distributions reflected rather faithfully the structure
of the classical repeller.

%

\begin{acknowledgments}

We acknowledge support from the Ministerio de Educaci{\'o}n y
Ciencia (Spain) [projects MTM2006-15533, MTM2009-14621 and CONSOLIDER
2006-32 (i-Math)]; the Comunidad de Madrid [project
S--0505/ESP-0158 (SIMUMAT)]; a grant (to SDP) from NanoforumEULA
 (support action funded by the European Union), and the NSF (USA) through grant CHE-0718547.
\end{acknowledgments}
%
\bibliography{GamowZA}%
\end{document}